\documentclass[sigconf]{acmart}
\AtBeginDocument{%
  }

\graphicspath{{figs/}{figures/}{pictures/}{images/}{./}} 

\usepackage{xcolor}
\usepackage{graphicx}
\usepackage{tikz}
\usepackage{svg}

\usepackage{amsmath,amsfonts}
\usepackage{algorithm}
\usepackage{algorithmicx}
\usepackage{algpseudocode}

\newcommand{\toolName}{\textit{Chameleon}}
\newcommand{\q}[1]{\textit{``#1''}}
\newcommand{\req}[1]{\textbf{\textit{#1}}}
\newcommand{\eg}{e.g.,\ }
\definecolor{custombox}{HTML}{FFF0DD}

\newcommand{\inlineimg}[1]{\raisebox{-0.25\height}{\includegraphics[height=1.7\fontcharht\font`\A]{#1}}}

\newcommand{\cref}[1]{Figure~\ref{#1}}

\setcopyright{acmlicensed}
\copyrightyear{2025}
\acmYear{2025}
\acmDOI{XXXXXXX.XXXXXXX}
\acmConference[ICHEC '25]{International Conference on Human-Engaged Computing}{November 21--23,
  2025}{Singapore, Singapore}
\begin{document}

\title{\toolName: Automated Color Palette Adaptation for Dark Mode Data Visualizations}

\author{Manusha Karunathilaka}
\authornote{Both authors contributed equally to this research.}
\affiliation{%
  \institution{Singapore Management University}
  \city{Singapore}
  \country{Singapore}
}
\email{gmik.vidana.2023@phdcs.smu.edu.sg}
\orcid{0009-0001-1345-0815}

\author{Songheng Zhang}
\authornotemark[1]
\affiliation{%
  \institution{Singapore Management University}
  \city{Singapore}
  \country{Singapore}
}
\email{shzhang.2021@phdcs.smu.edu.sg}
\orcid{0000-0002-0191-220X}

\author{Anthony Tang}
\affiliation{%
  \institution{Singapore Management University}
  \city{Singapore}
  \country{Singapore}
}
\email{tonyt@smu.edu.sg}
\orcid{0000-0003-4293-4082}

\author{Kotaro Hara}
\affiliation{%
  \institution{Singapore Management University}
  \city{Singapore}
  \country{Singapore}
}
\email{kotarohara@smu.edu.sg}
\orcid{0000-0002-7893-6090}

\author{Jiannan Li}
\affiliation{%
  \institution{Singapore Management University}
  \city{Singapore}
  \country{Singapore}
}
\email{jiannanli@smu.edu.sg}
\orcid{0000-0001-8409-4910}

\author{Yong Wang}
\authornote{Corresponding author.}
\affiliation{%
  \institution{Nanyang Technological University}
  \city{Singapore}
  \country{Singapore}}
\email{yong-wang@ntu.edu.sg}
\orcid{0000-0002-0092-0793}


\begin{abstract}
    Dark mode has gained widespread adoption across mobile platforms due to its benefits in reducing eye strain and conserving battery life. However, while the mobile system switches to dark mode, most visualizations remain designed for light mode, causing visual disruptions.
    Existing methods, such as manual adjustment or color inversion, are either time-consuming or fail to preserve the semantic meaning of colors in visualizations, making them less effective in dark mode. To address this challenge, we propose \toolName{}, an algorithm that automatically transforms light mode visualizations into dark mode while maintaining visual clarity and color semantics. By optimizing for luminance contrast, color consistency, and adjacent color differences, \toolName{} ensures that the transformed visualizations are legible and visually coherent. Our evaluation includes case study, expert interview, system evaluation, and a user study, and these demonstrate that \toolName{} is effective at translating visualizations for dark mode.
\end{abstract}





\begin{teaserfigure}
  \includegraphics[width=\textwidth]{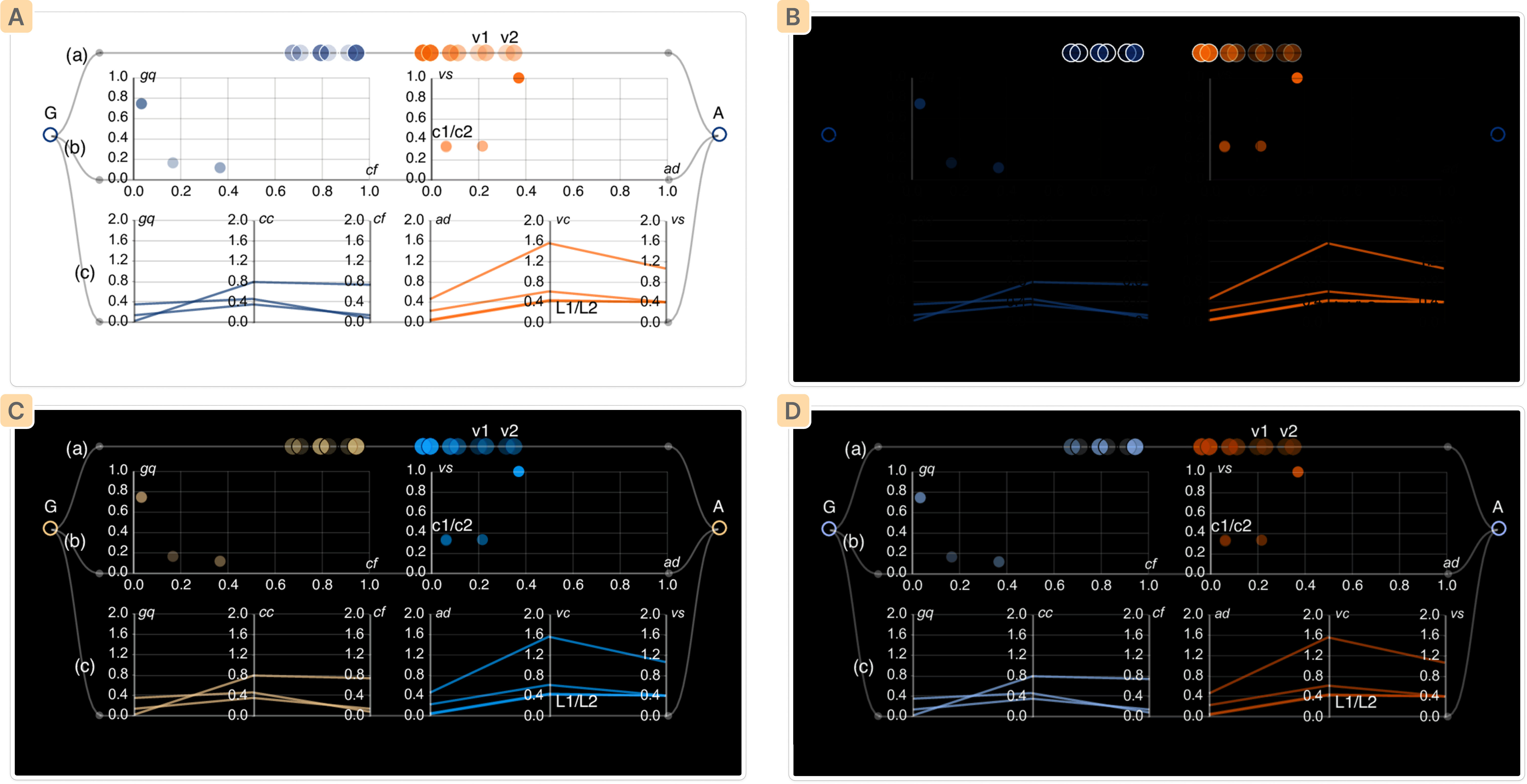}
  \caption{Visualization examples adapted from Shen et al.~\cite{Shen2017NameClarifierAV}, demonstrating four different rendering conditions. 
  (A) Original visualization in light mode. 
  (B) A typical dark mode conversion, where only the background is darkened while the original foreground colors are left unchanged, often leading to poor contrast and reduced legibility.
  (C) Visualization rendered using direct color inversion, which can distort the intended color semantics and lead to misinterpretation.
  (D) Visualization transformed using our method, \toolName{}, which optimizes color contrast and preserves semantic color mappings for improved legibility in dark mode.}
  \label{fig:teaser}
\end{teaserfigure}


\maketitle

\section{Introduction}

In recent years, dark mode has gained popularity across mobile platforms like Android and iOS due to its benefits, such as reduced eye strain in low-light conditions and prolonged battery life on OLED and AMOLED screens~\cite{Andrew2024LightAD,Iyer2003EnergyAdaptiveDS}. 
Dark mode is a user interface (UI) theme that uses a dark background with light text, as opposed to the default light mode, which typically uses a light background with dark text. 
Users can easily switch to dark mode through system settings, resulting in an immediate change in the system's UI. 
Visualizations are commonly used on mobile devices because of their convenience~\cite {Choe2019MobileDV, Lee2020ReachingBA, Roberts2014VisualizationBT}. 
When dark mode visualizations are effectively designed, they minimize distractions~\cite{appleDarkMode}, make important content stand out, and reduce eye strain~\cite{Qiao2023LightMA, Popelka2022TheEO}. 
They can also be made to adjust to changing light conditions, improving user comfort in dim environments~\cite{Qiao2023LightMA}. 
For instance, Mairena \textit{et al.}~\cite{Mairena2021WhichET} found that the users performed better using  the dark mode scatterplots, with faster search time and higher accuracy, compared to light mode.

However, most visualizations were designed for light mode, and when the system switches to dark mode, the mismatch between visualization elements and the dark background creates significant visibility problems.
\cref{fig:teaser}\inlineimg{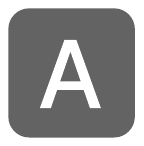} shows a visualization from a scientific paper~\cite{Shen2017NameClarifierAV} that is clearly visible against a light background. 
Yet, when the system switches to dark mode, as shown in \cref{fig:teaser}\inlineimg{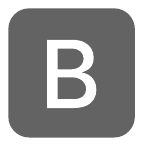}, the visualization elements become difficult to discern against the dark background, compromising their legibility. 
A common shortcut is to simply invert the color palette~\cite{Andrew2024LightAD}, as shown in \cref{fig:teaser}\inlineimg{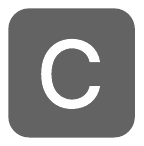}. 
Unfortunately, this inversion cannot preserve the color semantic relationships between the inverted visualization and the original visualization: the blue color in the original visualization is inverted to yellow, and the orange color is inverted to blue. 
This change largely disrupts users' cognitive ability to recognize and interpret the visualization across both modes.

As a consequence, designers are left to manually adapt visualizations for dark mode. 
This approach requires designers to build dark mode visualizations from scratch, a time-consuming process that involves complex multi-factor considerations: maintaining adequate contrast against the dark background, preserving the original color relationships, and ensuring overall visual harmony. 
These factors often conflict with each other. 
For example, increasing contrast for better legibility might disrupt color relationships or visual harmony. 
Since such factors are interdependent, designers must engage in extensive trial and error, making repeated adjustments to balance these competing requirements. 
While operating systems like Android and iOS offer guidelines~\cite{appleColorApple, appleDarkMode, androidDarkMode, pinterest} for UI design in dark mode, these guidelines typically focus on discrete UI elements like buttons and menus.
As a consequence, the guidelines are often too abstract to be directly applicable to the specific needs of visualizations~\cite{Andrew2024LightAD, Qiao2023LightMA}, which requires more nuanced design considerations. 

To reduce designers' workload, there is a need for an automated tool that can transform light mode visualizations to dark mode by adjusting their color palettes.
While there are existing tools for automatic color palette generation~\cite{Lu2020PalettailorDC, Gramazio2017ColorgoricalCD, Nardini2021AutomaticIO}, these cannot be directly applied to dark mode transformation.
These tools focus solely on generating optimized palettes for single mode and do not address the challenges of cross-mode transformation. 
These single-mode approaches creates several problems. 
First, colors that appear legible in light mode may become harsh or difficult to distinguish against dark backgrounds, an issue not considered by existing automated methods.
Second, colors need to maintain consistency between the modes to help users quickly interpret visualizations regardless of display mode. 
Finally, visual harmony and hierarchy must be preserved. 
For example, if highlighted regions have distinct contrast with adjacent areas in the light mode, this relationship should persist in the dark mode for consistent information emphasis.  
These cross-mode relationships and dark background interactions present unique challenges that require a different approach to the color palette generation.

To fill this gap, we propose \toolName{}, an algorithm designed to automatically transform visualizations from light mode to dark mode. 
Through literature reviews~\cite{appleDarkMode,androidDarkMode, Lu2020PalettailorDC, Gramazio2017ColorgoricalCD, Nardini2021AutomaticIO}, we identified three critical factors in dark mode visualization design: maintaining adequate luminance contrast for legibility, preserving color semantic relationships, and ensuring visual harmony between adjacent colors. 
The key challenge lies in quantifying these factors and balancing them effectively. 
\toolName{} addresses this by formulating the transformation as an optimization problem that mathematically models these design factors. 
Specifically, \toolName{} takes an image of a light mode visualization as input and analyzes its color palette. 
The algorithm then optimizes the dark mode color palette by balancing three designed loss functions: \textbf{\textit{luminance contrast consistency}} to ensure legibility against the dark background, \textbf{\textit{color consistency}} to maintain semantic relationships from the original visualization, and \textbf{\textit{adjacent color difference consistency}} to preserve visual harmony and hierarchy. 
This automated optimization eliminates the initial trial-and-error phase, providing designers with an optimized starting point that they can fine-tune rather than starting from scratch. 
The weights of these loss functions can be adjusted if designers want to prioritize different factors according to their specific needs. 
As shown in \cref{fig:teaser}\inlineimg{A.pdf} and \cref{fig:teaser}\inlineimg{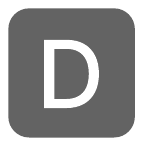}, this approach produces a dark mode visualization that successfully balances visual clarity, color consistency, and visual harmony.

To evaluate our approach, we conducted four evaluations. 
First, we produced several case study examples to demonstrate \toolName{}'s capability in handling various visualization types and color palettes. 
Next, we collaborated with professional visualization designers to understand how \toolName{} could streamline their dark mode transformation workflow. 
The expert interviews revealed that our method significantly reduced their workload by providing optimized starting points for further refinement. 
To investigate whether these automatically transformed visualizations could be used without designer intervention, we conducted a systematic evaluation against objective metrics for dark mode visualization design. 
Our results showed that the transformed visualizations met established standards for legibility and color consistency. 
Finally, to understand the practical impact of directly using those visualizations, we conducted a user study with 12 participants, examining how these transformed visualizations affected user preferences and task performance in real-world scenarios. 
We found that participants preferred \toolName{}'s results over other baselines, indicating its effectiveness in producing usable dark mode visualizations.

The major contributions of this paper can be summarized
as follows:
% \vspace{-1em}
\begin{itemize}
    \item \textbf{An algorithm}, \toolName{}, that automatically transforms light mode visualizations into dark mode by optimizing color palettes while balancing luminance contrast, color semantics, and relative perceptual differences.
    \item \textbf{A set of key design considerations} for dark mode visualization palettes, identified through literature reviews and expert interviews.
    \item \textbf{A comprehensive evaluation} of \toolName{} through a case study, expert interviews, system evaluation and a user study, demonstrating its effectiveness and usability.
\end{itemize}

\section{Related work}

\subsection{Color Palette Design Principles}\label{related_work:color_design_principle}

Color palette design for charts centers on two main considerations: perceptual discriminability and aesthetics~\cite{Yuan2021InfoColorizerIR}. 
The perceptual discriminability ensures that users can easily distinguish between colors within a visualization~\cite{Gramazio2017ColorgoricalCD, Stone2014AnEM}, a requirement that is critical for accurate data interpretation.
Healey~\cite{Healey1996ChoosingEC} emphasized that \q{colors should be well-separated,} underscoring the importance of maximizing visual distinctness.
Several factors influence discriminability: the data properties being represented (\eg continuous vs. categorical data)~\cite{Levkowitz1992ColorSF, Tennekes2014TreeCC}, the analytic tasks users perform (\eg comparing classes, identifying outliers)~\cite{Tominski2008TaskDrivenCC}, the ease of color naming and categorical labeling~\cite{Heer2012ColorNM}, and the degree of perceptual contrast among colors~\cite{Wang2019OptimizingCA, Mittelstdt2014MethodsFC}.
Healey~\cite{Healey1996ChoosingEC} proposed selecting representative colors from ten hue regions in Munsell’s color space while maximizing perceptual distances between them, while Maxwell~\cite{maxwell2000visualizing} similarly sought to maximize class discriminability, accounting for spatial distribution of colors.
However, these approaches face limitations like the resulting palettes are not always aesthetically pleasing~\cite{Lu2020PalettailorDC}, stressing that a proper balance between perceptual discriminability and aesthetics.

Beyond discriminability, palettes must also be visually appealing, as aesthetic qualities influence user engagement and interpretability.
Research shows that color harmony, the balanced arrangement of colors, enhances visual appeal and reduces cognitive strain~\cite{Meier2004InteractiveCP}.
Furthermore, manipulating hue, saturation, and brightness allows designers to balance vividness with readability, ensuring that palettes remain both expressive and coherent~\cite{Moreland2009DivergingCM}.
Well-chosen palettes thus balance clarity and beauty, encouraging users to focus on insights rather than being distracted by jarring color choices.
These design principles take on added importance in dark mode settings, where background inversion can drastically alter perceived contrast and harmony.
Guidelines from major platforms such as Android~\cite{androidDarkMode} and iOS~\cite{appleDarkMode} stress the importance of maintaining strong luminance contrast between visualization elements and background.
Without such care, a palette optimized for light mode may become unreadable, misleading, or aesthetically unbalanced in dark mode environments.
In this work, we build on these principles to address the unique challenges of dark mode visualization.

\subsection{Color Palette Generation}
The existing approaches to color palette generation fall into two categories: pre-defined and automatic.
Pre-defined palettes are carefully curated by experts, often based on principles of color theory and perceptual research.
A prominent example is ColorBrewer~\cite{Harrower2003ColorBrewerorgAO}, a widely adopted tool that provides palettes for encoding sequential (ordered values), diverging (centered around a critical midpoint), and qualitative (categorical classes) data.
These palettes are designed to maximize readability across print and digital media, and they have become a de facto standard in visualization practice. 
Similarly, visualization platforms such as Tableau provide pre-defined color palettes embedded into their systems~\cite{Milligan2018Tableau1C}, allowing designers to quickly apply professionally designed schemes.
While effective for common tasks, pre-defined palettes are inherently limited in flexibility. 
They typically support only a small set of colors and do not adapt to the specific characteristics of new datasets or visualization contexts, making them less suitable for complex visualizations requiring larger or interdependent color sets.

Automatic palette generation methods attempt to overcome these limitations by producing custom palettes that adapt to particular visualization needs.
These methods often employ optimization algorithms or learning-based approaches to balance perceptual discriminability and aesthetic preference.
For example, Colorgorical~\cite{Gramazio2017ColorgoricalCD} generates palettes by optimizing a user-defined balance between color discriminability and aesthetic preference for categorical data. 
The Palettailor~\cite{Lu2020PalettailorDC} incorporates data characteristics and employs a simulated annealing–based optimization when generating color palettes, assigning colors in a way that enhances visual discrimination between classes while preserving overall perceptual harmony.
While such approaches focus on generating palettes for categorical data, our method tackles the cross-mode adaptation problem: transforming palettes designed for light mode into ones suitable for dark mode.
To achieve this, we define the optimization problem around three dark mode–specific criteria, luminance contrast, semantic consistency, and relative color differences, ensuring that adapted palettes remain legible, meaningful, and visually coherent.

\section{Method}

\subsection{Dark Mode Visualization Transformation Factors}\label{method:transformtion_factors}

While there is extensive research on visualization design principles and color theory~\cite{schanda2007colorimetry, Healey1996ChoosingEC, Yuan2021InfoColorizerIR, Wang2019OptimizingCA}, and major platforms provide dark mode UI guidelines~\cite{appleDarkMode, androidDarkMode}, adapting visualizations specifically for dark mode remains unexplored. We analyzed dark mode UI design principles and identified three critical design factors as design requirements (DR) for effective dark mode visualization transformation:

\begin{itemize} 

\item \textbf{DR1: Perceptual Discriminability.} Visualization elements must remain clearly distinguishable against the dark background, just as they are in the light mode. 
Vision research further demonstrates that when luminance contrast decreases, color differences play a critical role in preserving discriminability~\cite{Chen2022ColorPolarity}.
This aligns with industry guidelines for dark mode from Android~\cite{androidDarkMode} and iOS~\cite{appleDarkMode}, and with visualization research emphasizing perceptual separability through optimized color assignment~\cite{Wang2019OptimizingCA}.

\item \textbf{DR2: Color Semantic Consistency.} Colors must maintain their semantic relationships between the light and the dark modes~\cite{Erickson2020DarkLightMA, Andrew2024LightAD}. 
For example, in visualization contexts where red indicates negative values and green indicates positive values, these color-meaning associations should remain consistent across modes. 
Schloss et al.~\cite{Schloss2019ColorMeaning} 
show that colormaps are effective when color–quantity mappings align with users’ expectations (\eg darker colors map to larger quantities).
Preserving these semantic relationships in dark mode is therefore essential to ensure consistency and interpretability.

 \item \textbf{DR3: Relative Perceptual Differences.} 
 The relative visual differences between adjacent colors must be preserved when transforming to dark mode. 
 For example, in a visualization where highlighted data points (e.g., outliers) are designed to stand out from regular data points through color contrast, this visual emphasis should remain equally noticeable after transformation~\cite{Lu2023InteractiveCC}. 
 If a bright orange highlight distinctly contrasts with blue regular points in light mode, their transformed colors in dark mode should maintain this level of differentiation.
 Preserving this relative color differences ensure that visual hierarchy and emphasis techniques remain effective across modes~\cite{Mittelstdt2014MethodsFC}.

\end{itemize}

\subsection{Problem Formulation}\label{method:problem_formulation}

\toolName{} takes as input a bitmap image of a light-mode visualization, where each pixel is represented by an RGB color value.
To process the visualization, 
users need to specify three parameters: 
(1) the light mode background color used in the input visualization (typically white but can be any light color),
(2) the target dark mode background color 
(typically black or dark gray), 
and 
(3) the number $k$ of distinct colors to extract the visualization's color palette. 
Given these parameters, we first obtain the visualization's color palette (detailed in Section~\ref{method:color_palette_extraction}). 
We then adapt the obtained color palette to a new color palette, which is suitable for the dark mode visualization. 

To make this adaptation optimal, we formulate the dark mode palette generation as an optimization problem. 
The objective is to derive a palette $P=\{c_1, \dots, c_m\}$,
where each $c_i$ represents a unique color, such that the palette simultaneously preserves the semantic meaning of the original colors and maintains the visual clarity against the dark background.  
Formally, the optimization seeks the palette 
$P$ that minimizes the following objective function:

\begin{equation}
\text{argmin}_P \, E(P) = \omega_0 E_{LC} + \omega_1 E_{CC} + \omega_2 E_{AC}.
\label{eq:object_function}
\end{equation}

The object function ($E(P)$) is a weighted sum of three loss functions: \underline{L}uminance \underline{C}ontrast Consistency from \req{DR1} ($E_{LC}$) , \underline{C}olor \underline{C}onsistency from \req{DR2} ($E_{CC}$), and \underline{A}djacent \underline{C}olor Difference Consistency from \req{DR3} ($E_{AC}$). 
% Each color $c_i$ in the color palette represents a unique color, and 
The weights $\omega_0$, $\omega_1$, and $\omega_2$ control the relative importance of each criterion, allowing a balance between preserving semantic meaning and ensuring legibility in the dark mode.
The next subsections detail these loss functions.

\subsection{Color Palette Extraction
}\label{method:color_palette_extraction}

\begin{figure}
    \centering
    \includegraphics[width=1\linewidth]{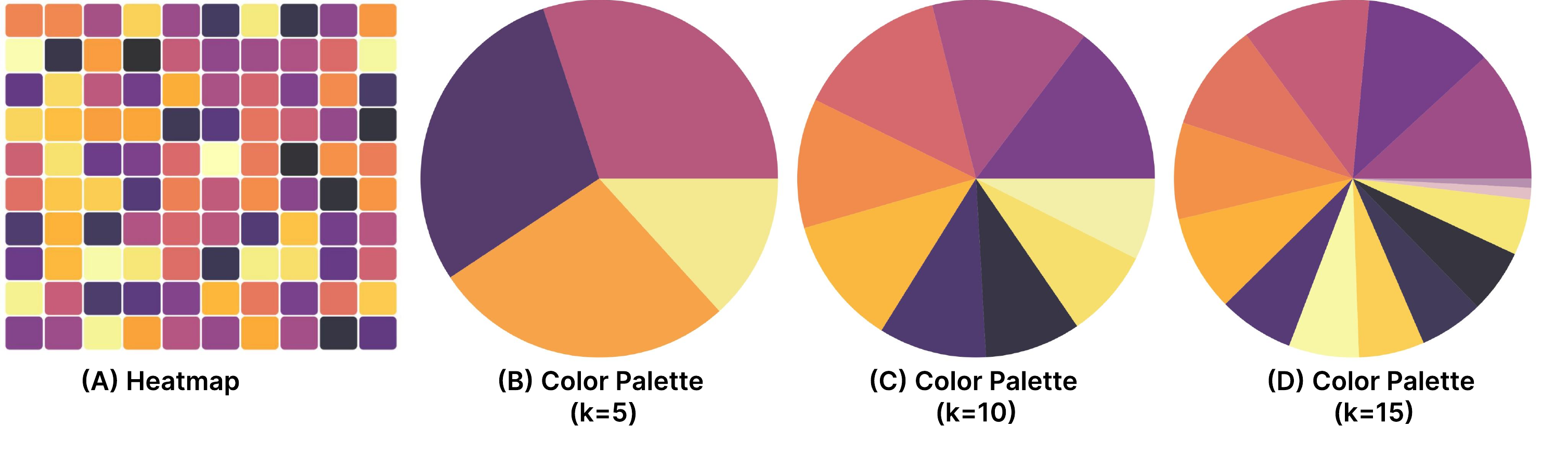}
    \caption{An illustration of how k-means clustering with varying $k$ values affects the extracted color palettes from a visualization. (A) shows a heatmap visualization. (B), (C), and (D) represent the identified colors after clustering with $k$ values of 5, 10, and 15, respectively. As the value of $k$ increases, more distinct color clusters are identified from the visualization, capturing finer details and variations in the color palette. }
    \label{fig:kmean_example}
\end{figure}

\begin{figure}
    \centering
    \includegraphics[width=1\linewidth]{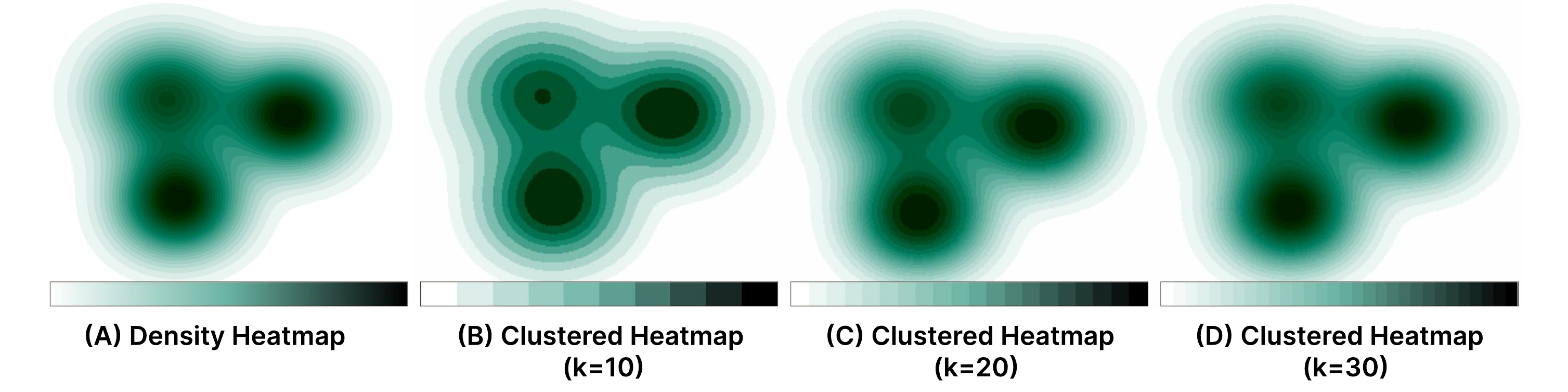}
    \caption{An illustration of how k-means clustering with different $k$ values affects a continuous visualization. (A) shows a density heatmap visualization. (B), (C), and (D) represent the density heatmap clustered using k-means with $k$ values of 10, 20, 30, respectively. The bottom legend shows its clustered $k$ colors. Increasing the value of $k$, resulting in a finer transition between colors within the density heatmap.}
    \label{fig:eg_continuous_heatmap}
\end{figure}

To extract the color palette from the input image, \toolName{} uses k-means clustering~\cite{Hartigan1979AKC} on the RGB values of all non-background pixels, where each pixel is treated as a three-dimensional point (R, G, B).
The clustering process groups similar colored pixels together and identifies $k$ cluster centroids, which form our extracted color palette.
As shown in~\cref{fig:kmean_example}, the choice of $k$ affects the granularity of color identification, where a larger $k$ value captures more distinct colors and subtle variations in the visualization. 
For example, when $k$=5, the algorithm identifies only the most dominant colors, while $k$=15 captures more nuanced color variations. 
For continuous visualizations (\cref{fig:eg_continuous_heatmap}), such as heatmaps or density plots, the impact of $k$ becomes even more significant.
A higher number of clusters ensures smoother transitions between colors, as demonstrated in \cref{fig:eg_continuous_heatmap}, where $k$=30 produces smoother gradients than $k$=10.
This flexibility in $k$ selection allows \toolName{} to handle both visualizations with distinct color categories and those requiring smooth color transitions.

\subsection{Loss Functions and Design}\label{method:loss_functions_designs}

We operationalize three concepts, luminance contrast, color consistency, and adjacent color difference through three loss functions.

\textbf{Luminance Contrast Consistency:} 
Luminance contrast plays a critical role in object-background differentiation~\cite{Zhang2023DontPA, Lubin1997AHV}. 
A sufficient level of luminance contrast enables the human eye to distinguish the visual elements from the background, thereby ensuring the clarity of the visualization.
In our work, the function aims to maintain the luminance contrast between visualization elements with the dark mode background is comparable to that of the light mode background. 
This allows the human eye to perceive the same contrast of elements in the dark mode background as over the light mode background, thereby maintaining clarity across different modes.
In our approach, we utilize the L channel from the LCH color space~\cite{schanda2007colorimetry}, as it closely matches human perception of luminance~\cite{Hanbury2011MATHEMATICALMI}.

The luminance contrast consistency loss is calculated as follows:
\begin{equation}
E_{\text{LC}} = \left| |L_{\text{bg}}^{\text{light}} - L_i^{\text{light}}| - |L_i^{\text{dark}} - L_{\text{bg}}^{\text{dark}}| \right|
\label{eq:luminance_contrast_consistency},
\end{equation}

where $L_{\text{bg}}^{\text{light}}$ and $L_i^{\text{light}}$ represent the lightness values for the light mode background and a specific color $i$, respectively, while $L_{\text{bg}}^{\text{dark}}$ and $L_i^{\text{dark}}$ represent the corresponding values in the dark mode.
This formulation ensures that luminance contrast between elements and background remains comparable across modes, keeping visualization elements perceptually discriminable (\req{DR1}).

\textbf{Color Consistency}: 
In many visualizations, color choices carry semantic meaning (\eg ~\cref{fig:teaser}). 
Maintaining color consistency helps users to interpret the data in both light and dark modes without confusion. 
This function aims to preserve the semantic meaning relationships 
of colors by ensuring that the colors in the dark mode remain perceptually similar to those in the light mode. 
We evaluate color consistency using the Delta CIE2000 metric~\cite{Sharma2005TheCC}, which measures perceptual color differences:

\begin{equation} 
E_{\text{CC}} = \left| \Delta E_{2000}(C_{i}^{\text{light}}, C_{i}^{\text{dark}}) \right|, \label{eq:color_consistency}
\end{equation}

where \( C_{i}^{\text{light}} \) represents a specific color in the original light mode visualization, and \( C_{i}^{\text{dark}} \) represents its corresponding transformed color in the dark mode.
$\Delta E_{2000}$ represents the color difference between the light mode and dark mode color $C_i$. 
The $\Delta E_{2000}$ metric was chosen because it better matches human perception of color difference compared to earlier color difference formulas like CIELAB ($\Delta E_{ab}$) or Delta CIE94 ($\Delta E_{94}$)~\cite{Sharma2005TheCC}.
This formulation ensures that transformed colors remain perceptually close to their originals, preserving semantic color meaning across modes (\req{DR2}).

\textbf{Adjacent Color Difference Consistency:}
Maintaining the relative differences between colors in the visualization helps users distinguish between different data elements, ensuring harmony in the transformed visualization. This function ensures that the differences between adjacent colors remain consistent between light and dark modes. 
In a visualization, a color's adjacent colors are those that appear in direct spatial proximity to it. 
These can be identified by examining the immediate neighbors of any colored element: for example, in a heatmap, each cell's adjacent colors are those of its neighboring cells in all directions. 
Preserving these relative color differences is crucial for maintaining the visualization's visual hierarchy and relationships. 
We compute adjacent color differences at the cluster level: each color cluster is represented by its CIELAB centroid, and the loss is calculated as follows:
% The adjacent color difference consistency loss is calculated as follows:

\begin{equation}
\begin{aligned}
E_{\text{AC}} = \frac{1}{|\mathcal{A}|} \sum_{a \in \mathcal{A}} \Big| \Delta E_{2000}(C_{i}^{\text{light}}, C_{a}^{\text{light}}) 
- \Delta E_{2000}(C_{i}^{\text{dark}}, C_{a}^{\text{dark}}) \Big|,
\end{aligned}
\end{equation}

where $ \mathcal{A} $ is the set of colors adjacent to color $ C_{i} $, and $ \Delta E_{2000} $ calculates the perceptual color difference between color $ C_{i} $ and its adjacent colors $ C_{a} $, while $C^{\text{light}}_i$ and $C^{\text{light}}_a$ denote the light mode colors and $C^{\text{dark}}_i$ and $C^{\text{dark}}_a$ are their dark mode counterparts.
This formulation penalizes deviations in perceptual differences between $C_i$ and its adjacent colors $C_a$, thereby preserving their relative distinguishability across modes (\req{DR3}).

\subsection{Simulated-Annealing-based Optimization}\label{method:algorithm_optimization_process}

The optimization process follows a simulated annealing algorithm (Algorithm~\ref{table:algorithm}) to find the best color palette. 
Given the vast solution space with numerous possible color combinations, simulated annealing is well-suited for exploring such large spaces efficiently, balancing between exploration (trying new palettes) and exploitation (refining towards the best solution).
Furthermore, we choose simulated annealing over learning-based or heuristic alternatives because it provides explicit, interpretable control over perceptual loss functions without requiring training data. 
This keeps the optimization transparent, reproducible (given seeds), and directly governed by our defined perceptual metrics rather than learned priors.
Starting with an initial color palette $P_0$, the algorithm iterates through possible palettes, and refines them to find an optimal solution $P^*$.

\textbf{Initialization (Lines 1–5).}
The algorithm initializes three variables: the initial color palette $P_0$, the current palette $P_{\text{current}}$, and the optimized palette $P^*$. 
At the start, both $P_{\text{current}}$ and $P^*$ are set to $P_0$.
The $P_0$ initializes with the color palette extracted from the input visualization via k-means clustering (Section~\ref{method:color_palette_extraction}).
Furthermore, the algorithm sets parameters, including the initial temperature $T_0$=10,000 and a cooling rate $\alpha$=0.99, to guide the exploration of the solution space over 20,000 iterations.

\begin{algorithm}
\caption{Simulated Annealing Algorithm}
\begin{algorithmic}[1]
\State \textbf{Input:} Initial color palette $P_0$, initial temperature $T_0$, cooling rate $\alpha$, number of iterations $N$
\State \textbf{Output:} Optimized color palette $P^*$
\State Initialize $P^* \leftarrow P_0$
\State Initialize $P_{\text{current}} \leftarrow P_0$
\State Initialize $T \leftarrow T_0$

\For{$i = 1$ to $N$}
    \State $P_{\text{new}} \leftarrow$ Randomly disturb one color from $P_{\text{current}}$
    \State $E_{\text{current}} \leftarrow E(P_{\text{current}})$
    \State $E_{\text{new}} \leftarrow E(P_{\text{new}})$

    \If{$E_{\text{new}} < E_{\text{current}}$}
        \State $P_{\text{current}} \leftarrow P_{\text{new}}$
        \If{$E_{\text{new}} < E(P^*)$}
            \State $P^* \leftarrow P_{\text{new}}$
        \EndIf
    \Else
        \State $\Delta E \leftarrow E_{\text{new}} - E_{\text{current}}$
        \State $p \leftarrow \exp\left(-\frac{\Delta E}{T}\right)$
        \If{random(0, 1) $< p$}
            \State $P_{\text{current}} \leftarrow P_{\text{new}}$
        \EndIf
    \EndIf

    \State $T \leftarrow \alpha T$
\EndFor

\State \Return $P^*$
\end{algorithmic}
\label{table:algorithm}
\end{algorithm}

\textbf{Generating a new color palette (Lines 7–9).}
At each iteration, the algorithm generates a new palette $P_{\text{new}}$ by randomly adjusting one color in $P_{\text{current}}$. 
We operate in the LCH (Lightness, Chroma, Hue) color space, which provides more intuitive control over color properties than RGB~\cite{schanda2007colorimetry}. 
For each adjustment, we randomly select one of three color components with equal probability: lightness (L), chroma (C), or hue (H). 
If lightness is selected, we adjust its value by a random amount between -20 and +20 units and ensure the result stays within [0, 100], where 0 represents black and 100 represents white. 
For chroma adjustments, we similarly apply a random change between -20 and +20 units, constrained to [0, 100], where higher values indicate more color saturation. 
If hue is selected, we modify it by a random value between -50 and +50 degrees and constrain it to [0, 360], representing the full color wheel. 
Since not all colors in LCH color space can be represented on digital displays that use sRGB color space, we verify that each perturbed color falls within the valid sRGB gamut by checking whether its transformed RGB values lie between 0 and 1. 
If a perturbed color is outside this gamut, the algorithm continues generating new perturbations until a valid, displayable color is found or a maximum attempt limit is reached. 
This controlled random perturbation strategy helps the algorithm explore the color space effectively while ensuring all generated colors can be correctly displayed.

\textbf{Finding the best color palette (Lines 10-24)}. 
The algorithm evaluates both $P_{\text{current}}$ and $P_{\text{new}}$ using the objective function. 
If the new palette $P_{\text{new}}$ decreases the loss, it is accepted as the current palette. 
The algorithm also probabilistically accepts worse solutions to avoid local optima, reducing this probability as the temperature cools. 
This iterative process continues until the optimal palette $P^*$ is found.

\section{Evaluation}
We assess the effectiveness of \toolName{} through four complementary studies.
First, we present case studies that demonstrate how \toolName{} adapts a variety of visualization types to dark mode. 
Second, we report insights from expert interviews with visualization and UI designers on its practical utility. 
Third, we conduct a systematic evaluation of legibility and color consistency using established quantitative metrics. 
Finally, we describe a user study that examines how \toolName{} impacts user perception, task performance, and visual fatigue.

\subsection{Case Studies of Visualizations}

\begin{figure}
    \centering
    \includegraphics[width=1\linewidth]{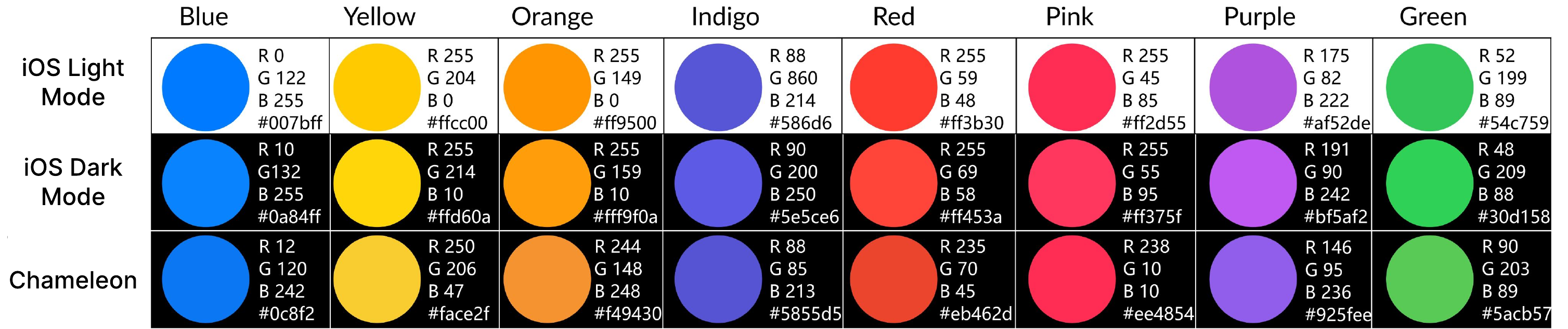}
    \caption{Comparison of color categories across three conditions: iOS light mode colors, iOS dark mode colors, and the corresponding dark mode colors generated by \toolName{} from the iOS light mode palette.}
    \label{fig:case_study_apple}
\end{figure}

We illustrate how the \toolName{} performs in comparison to several real-world cases. 
We demonstrate that \toolName{} generates color palettes that are comparable to manually defined palettes and can scale larger color palettes (Section~\ref{case:colors}). 
Further, we have found that \toolName{} generates palettes that are perceptually similar to those found in dark mode design guidelines (Section~\ref{case:simple_chart}). 
Finally, we demonstrate \toolName{} 's versatility across different visualization formats, including visualizations with continuous colormaps (e.g., temperature data), multiple-view visualizations, and infographics (Section~\ref{case:complex_chart}).

\subsubsection{\textbf{Comparisons on Color Palettes}}\label{case:colors}

\toolName{}'s automated transformations closely resemble Apple's predefined dark mode palette. 
Apple's iOS design guidelines~\cite{appleColorApple} offers a small set of predefined pairs of colors for light and dark modes, where each light mode color corresponds to a specific dark mode counterpart (e.g., light red becomes dark red). 
Using Apple's predefined light mode colors (first row in \cref{fig:case_study_apple}) as input, we applied \toolName{} to transform these colors into dark mode equivalents. 
Each light mode color (\eg \inlineimg{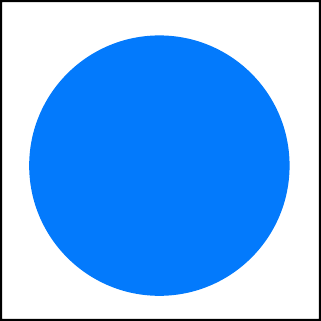}) was processed through our algorithm, generating its corresponding dark mode version (\eg \inlineimg{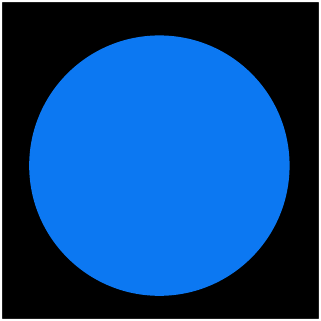}). 
We tested eight different colors, and the transformed dark mode colors are displayed in \cref{fig:case_study_apple}. 
From the top-to-bottom comparison in the last row of \cref{fig:case_study_apple}, we see that the dark mode colors generated by \toolName{} closely match the predefined dark mode colors from Apple (the second row of \cref{fig:case_study_apple}). 
However, the iOS predefined color palette is limited in the number of color pairs available, making it less adaptable for visualizations with more color requirements. 
In contrast, \toolName{}’s automated transformation process allows it to scale to a larger variety of colors, enabling use across visualizations that may exceed the limited set provided by iOS.

\subsubsection{\textbf{Visualization Examples from Dark Mode Guideline Documentation}}\label{case:simple_chart} 

\begin{figure}
    \centering
    \includegraphics[width=0.8\linewidth]{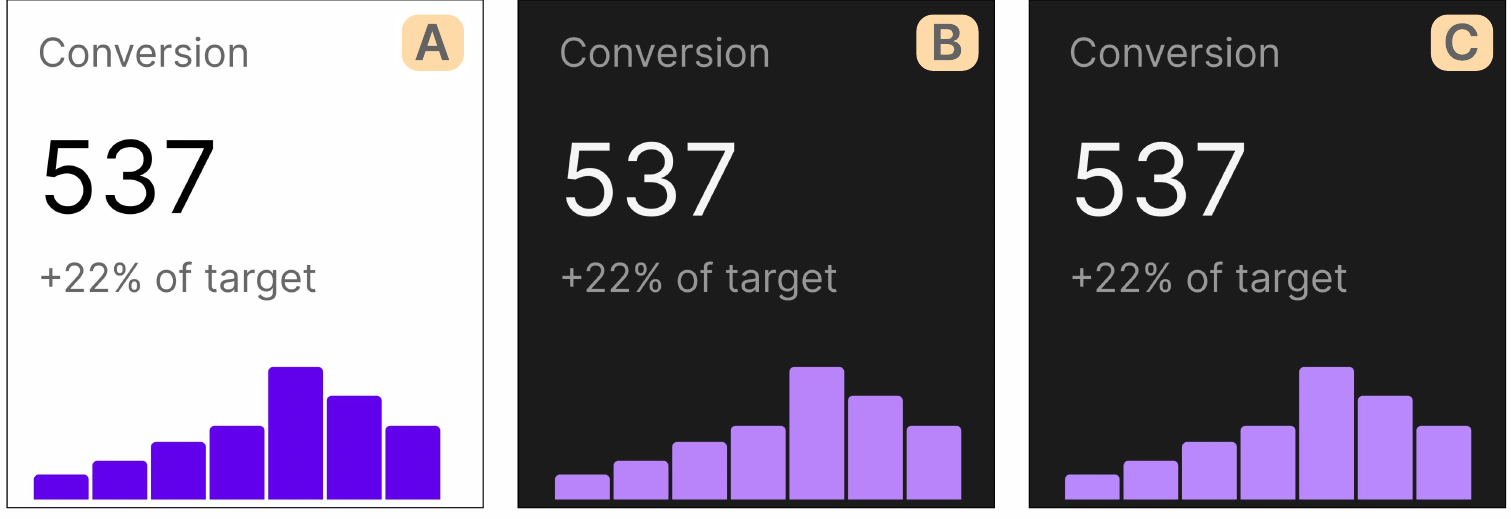}
    \caption{The comparison of visualizations in three conditions: (A) light mode visualization from the Android official design guidelines\protect\footnote{https://m2.material.io/design/color/dark-theme.html}
, (B) corresponding dark mode visualization from the guidelines, and (C) dark mode visualization transformed by \toolName{}.}       \label{fig:case_study_andorid}
\end{figure}

Our method transforms and produces results that align with the official dark mode design guidelines, such as those provided by Android. As seen in \cref{fig:case_study_andorid}\inlineimg{A.pdf} and \cref{fig:case_study_andorid}\inlineimg{B.pdf}, the Android dark mode design document~\cite{androidDarkMode} includes a pair of visualization examples for light and dark modes. 
Comparing \cref{fig:case_study_andorid}\inlineimg{B.pdf} (official dark mode example) and \cref{fig:case_study_andorid}\inlineimg{C.pdf} (output from \toolName{}), we observe similarities. 
Notably, the colors of the bar chart in our dark mode transformation closely match those provided in the official Android design. 
Additionally, text elements were automatically converted from black to white to maintain legibility against the dark background.

\subsubsection{\textbf{Applications to Different Visualization Formats}}\label{case:complex_chart}
\begin{figure}
    \centering
    \includegraphics[width=1\linewidth]{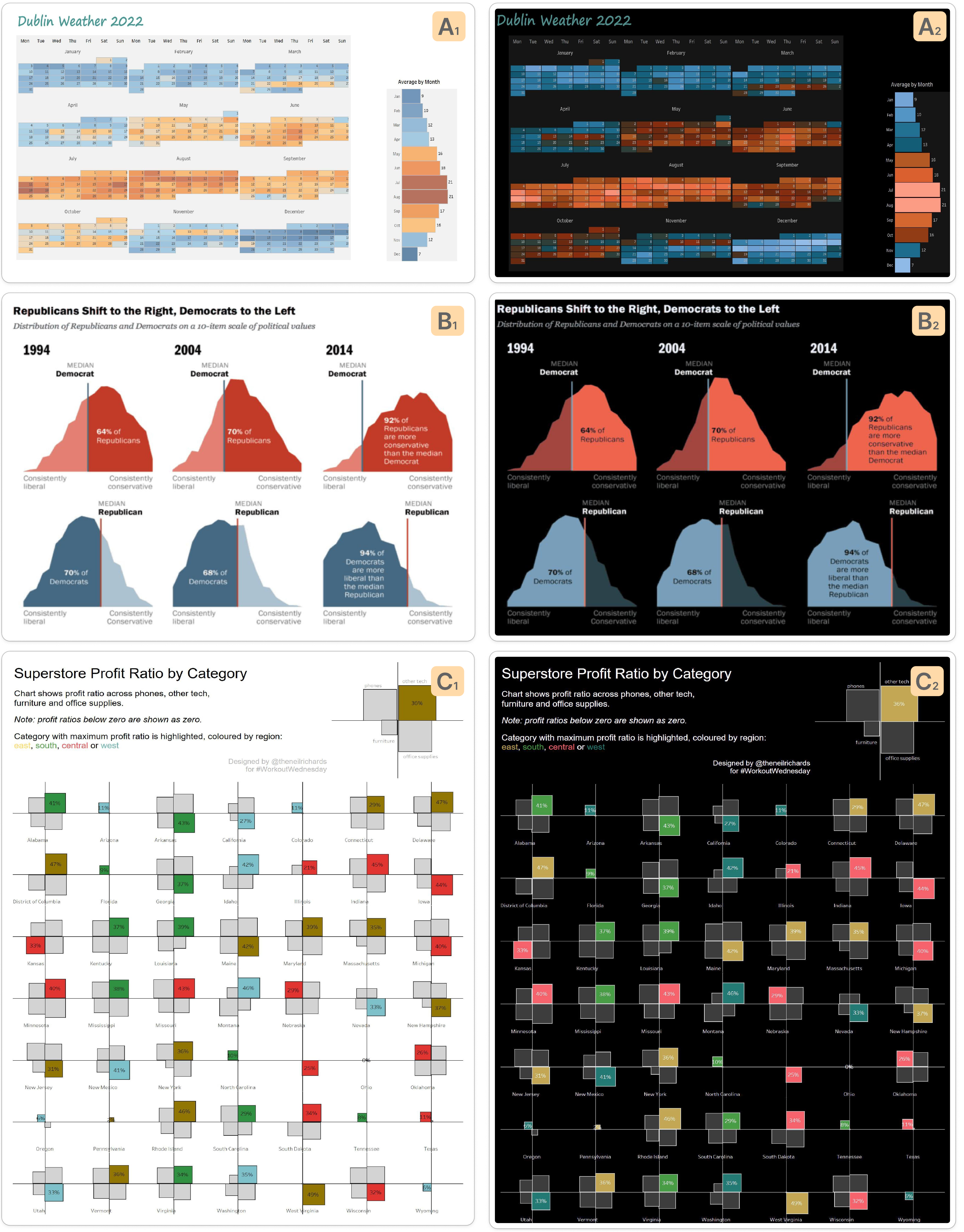}
    \caption{Examples of visualizations in the light mode and the corresponding dark mode results from \toolName{}. The left-side visualizations show the original ones in light mode, while the right-side visualizations are transformed by our method for dark mode.}
    \label{fig:case_study_complex}
\end{figure}

To evaluate \toolName{}'s effectiveness, we tested it on visualizations with varying formats and color patterns, focusing on three representative examples shown in \cref{fig:case_study_complex}.
The temperature visualization in \cref{fig:case_study_complex}\inlineimg{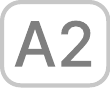} demonstrates \toolName{}'s ability to handle continuous colormaps. 
The original visualization (\cref{fig:case_study_complex}\inlineimg{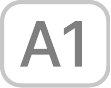}) uses a temperature scale from dark blue to dark red, which would be illegible on a dark background. 
\toolName{} successfully maintains luminance contrast (\req{DR1}) while preserving the smooth temperature gradient through consistent adjacent color differences (\req{DR3}).
\cref{fig:case_study_complex}\inlineimg{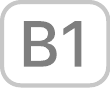} shows a political visualization that uses red-blue encoding to represent Democratic (blue) and Republican (red) political values, with highlighted areas of interest. 
\toolName{}'s result in \cref{fig:case_study_complex}\inlineimg{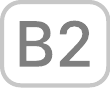} preserves both the semantic meaning of colors (\req{DR2}) and the visual hierarchy created by contrast differences (\req{DR3}), while maintaining readability against the dark background (\req{DR1}).
The profit ratio visualization in \cref{fig:case_study_complex}\inlineimg{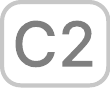} shows \toolName{}'s handling of categorical data using four distinct colors. 
\toolName{} preserves both color consistency (\req{DR2}) and adjacent color differences (\req{DR3}) while ensuring sufficient contrast with the dark background (\req{DR1}). This enables viewers to distinguish between categories as effectively as in the light mode version.

Based on these examples (among others in Appendix), we conclude that our approach can generate dark mode visualizations that maintain legibility, and maintain color semantics.

\subsection{Expert Interview}\label{expert_interview}

\begin{table}[t]
\footnotesize
\setlength{\tabcolsep}{4pt}
\begin{tabular}{@{}p{0.07\columnwidth}p{0.13\columnwidth}p{0.35\columnwidth}p{0.25\columnwidth}p{0.1\columnwidth}@{}}
\toprule
\textbf{ID} & \textbf{Gender} & \textbf{Job} & \textbf{Domain} & \textbf{Years} \\ 
\midrule
E1 & Female & Researcher & Visualization & 5 \\
E2 & Female & Researcher & Visualization & 4 \\
E3 & Male & Visual \& Data Journalist & Journalism & 2 \\
E4 & Female & UI/UX Design & Software & 2 \\
\bottomrule
\end{tabular}
\caption{Experts' demographic information.}
\label{table:exp_demo_info}
\end{table}

To understand how professionals would use a tool like \toolName{} in their workflow, and how they would perceive its performance, we conducted semi-structured interviews with four visualization design and UI design professionals. 
As shown in Table~\ref{table:exp_demo_info}, two of the experts (E1 and E2) have over four years of experience in visualization design research. 
E1 specializes in visual analysis system design, while E2 focuses on color design in visualizations. 
The other two experts (E3 and E4) have over one year of industry experience, with E3 working as a visualization designer for a newspaper and E4 as a UI/UX designer. 
Each interview lasted approximately 60 minutes and began with a brief introduction of the project. 
We then presented several cases for discussion, allowing participants to share their insights and comments. 
The interviews were recorded and transcribed using automated speech-to-text technology~\cite{Radford2022RobustSR}, followed by a thematic analysis to summarize their feedback.

\paragraph*{\textbf{Existing Dark Mode Design Workflow is Time-Consuming.}} 
All experts agreed that designing visualizations for dark mode is an iterative, trial-and-error process. 
They emphasized that transforming a visualization or UI from light to dark mode requires careful adjustments and refinements. 
Most experts start by changing the background to black and then assess the existing color palettes to ensure all visualization elements remain clear and accessible. 
Adjustments are made as needed to preserve legibility and information integrity. 
This is followed by iteratively fine-tuning visualization elements' details such as color contrasts to achieve a visually harmonious result. 
As E2 described, \q{When designing dark mode, the first thing I do is set the background to black and the text to white. I then look at the visualization without modifying the foreground colors, only changing the background, and assess how it looks. If I find certain areas that don't look right, I'll manually adjust those areas. I try to maintain the original elements (color palette) as much as adjustments, but adjust where necessary. I typically adjust colors gradually, tweaking the saturation and lightness, especially, if color palette includes high-saturation or dark colors that don't work well in dark mode. I ensure aesthetic harmony by keeping the color saturation and lightness consistent. For example, I avoid using colors that are too saturated or uncommon.} 
E2 also talked about her reluctance to use color inversion, \q{Rather than inverting everything (in the visualization), I prefer making adjustments in specific areas to maintain the integrity of the design.}

Transforming light mode visualizations to dark mode is time-intensive, particularly for complex designs like multiple-view visualizations. 
The experts highlighted the significant effort required to ensure the final dark mode design is both visually coherent and effective. 
E3 noted, \q{Designing a dark mode visualization definitely requires more work, often taking an additional few hours to a couple of days.} 
E2 further explained the complexity, especially when considering color harmony: \q{In light mode, colors are harmonious because their brightness or saturation is consistent. When transforming to dark mode, I have to preserve the harmony. For example, I cannot only adjust certain colors, because adjusting a single color can disrupt this balance. This makes dark mode design especially challenging.} 
E1 echoed this sentiment, emphasizing the importance of aesthetics and color harmony, estimating that the process could take up to two or three days.

\paragraph*{\textbf{\toolName{} Benefits.}} The experts felt that \toolName{} would reduce the workload and provide a strong foundation for further refinement. This is valuable, as manual adjustments for dark mode can be time-consuming and labor-intensive. 
E1 praised the method’s efficiency, stating, \q{With \toolName{}, I can complete the transformation much faster, as it handles much of the groundwork. I now only need to make minor adjustments compared to designing from scratch.} 
E1 also noted the psychological benefit, explaining that \toolName{} significantly reduced the mental burden of designing for dark mode: \q{It feels like the task has become more achievable, with a solid foundation already in place.} 
The experts agreed that \toolName{} provides a practical starting point that reduces the overall workload. 
E4 emphasized that our method has satisfied the design requirement, \q{When we began designing a user interface (UI), we established specific design requirements for the UI. [When designing a UI in dark mode], our goal was to ensure that the information in the dark mode UI meets the same design requirements as the light mode. The examples displayed in dark mode retain the same design requirements as the original visualizations shown in light mode.
In the figure (\cref{fig:case_study_complex}\inlineimg{C.pdf}), certain parts of the visualizations are highlighted. In the original visualization, these parts are immediately noticeable at a glance. In dark mode, these same parts are also the first things I noticed.} 
E2 added that the method successfully preserves the thematic colors of visualizations, stating, \q{Each dashboard usually has a primary color representing the theme. Your method keeps this primary color intact when transforming visualizations to dark mode.}

\paragraph*{\textbf{Opportunities for Improving Legibility with \toolName{}.}} 
A primary concern was the insufficient contrast of certain elements like axes and text, which are critical for understanding the visualization. 
E1 pointed out, \q{The legend and axis labels are too blurry, and I can’t clearly see the information. These elements need more contrast against the background.} 
E4 agreed, explaining, \q{Text and axes require higher brightness on a black background, a general rule in UI/UX design: you need higher brightness and saturation on a black background than on a white background.} 
To address this, E2 suggested treating axes and text as separate elements from the visualization and simply inverting their colors, stating, \q{In most cases, inverting text color from black to white is effective. However, more nuanced adjustments may be needed when the original text has color.} 
E2 also raised concerns about the aesthetic quality of the dark mode palette, particularly with certain colors appearing less visually pleasing.
\q{Some colors in the dark mode visualization, like yellow and green, aren’t as saturated or `clean' as I would expect.} 
E2 remarked, suggesting that rules could be applied to refine these colors.

A promising direction for future work is to apply our algorithm selectively to visualization components, focusing on color-encoded elements while semantically distinguishing them from meta-elements such as text and axes. 
We leave this extension for future work, as separating visual elements from meta-elements was beyond our current design scope.
\subsection{System Evaluation}

While \toolName{} primarily serves as a design aid, we evaluated whether its automatically transformed visualizations could be usable without designer intervention. 
We focused on quantitatively assessing legibility (\req{DR1}) and color semantic consistency (\req{DR2}). 
Visual hierarchy maintenance (\req{DR3}) was excluded from this evaluation due to its context-dependent nature and was instead assessed through expert interviews and case studies.

\subsubsection{Evaluation Design}

We evaluated 85 charts from the D3 Graph Gallery~\cite{d3graphgalleryGraphGallery}, spanning 37 distinct visualization types from basic (bar, line, pie charts) to complex forms (streamgraphs, choropleth maps). 
This diverse sample allowed us to test our method across varied visualization types and color palettes.
We compared three conditions:

\begin{itemize}
    \item \textbf{Light Mode}: Original visualizations with white background, serving as the baseline
    \item  \textbf{Inverse Mode}: Colors inverted by subtracting RGB values from 255, representing a common automated approach
    \item  \textbf{Dark Mode}: Colors transformed by our algorithm for dark background
\end{itemize}
We excluded original colors on black backgrounds (insufficient contrast as \cref{fig:teaser}\inlineimg{B.pdf}) and manual designer palettes (not automated method).

\subsubsection{Metrics and Procedures}
We assessed two key metrics:
Legibility measured the contrast ratio between visual elements and background according to WCAG 2.1 AA guidelines~\cite{w3ContentAccessibility}, requiring a minimum 3:1 ratio:
\begin{equation} \text{CR}(L_{1}, L_{2}) = \frac{L_{1} + 0.05}{L_{2} + 0.05}, \end{equation}

\begin{equation} \text{Contrast Ratio Threshold} = \frac{\sum_{i=1}^{n} P_i \times CR(C_{i}, C_{BG})}{P_{total}} \ge 3.0,
\label{eq:contrast_ratio_threshold}
\end{equation}

Color Consistency quantified the perceptual difference between original and transformed colors using the $\Delta E_{2000}$ metric:
\begin{equation}
    \text{Color Difference} = \frac{\sum_{i=1}^{n} P_i \times \Delta E_{2000}(C_{i, Light}, C_{i, Dark\backslash Inverse})}{P_{total}},
\label{eq:test_color_consistency}
\end{equation}

We extracted color palettes using k-means clustering and evaluated each visualization against these metrics.

\subsubsection{Results}
\begin{figure}
    \centering
    \includegraphics[width=1\linewidth]{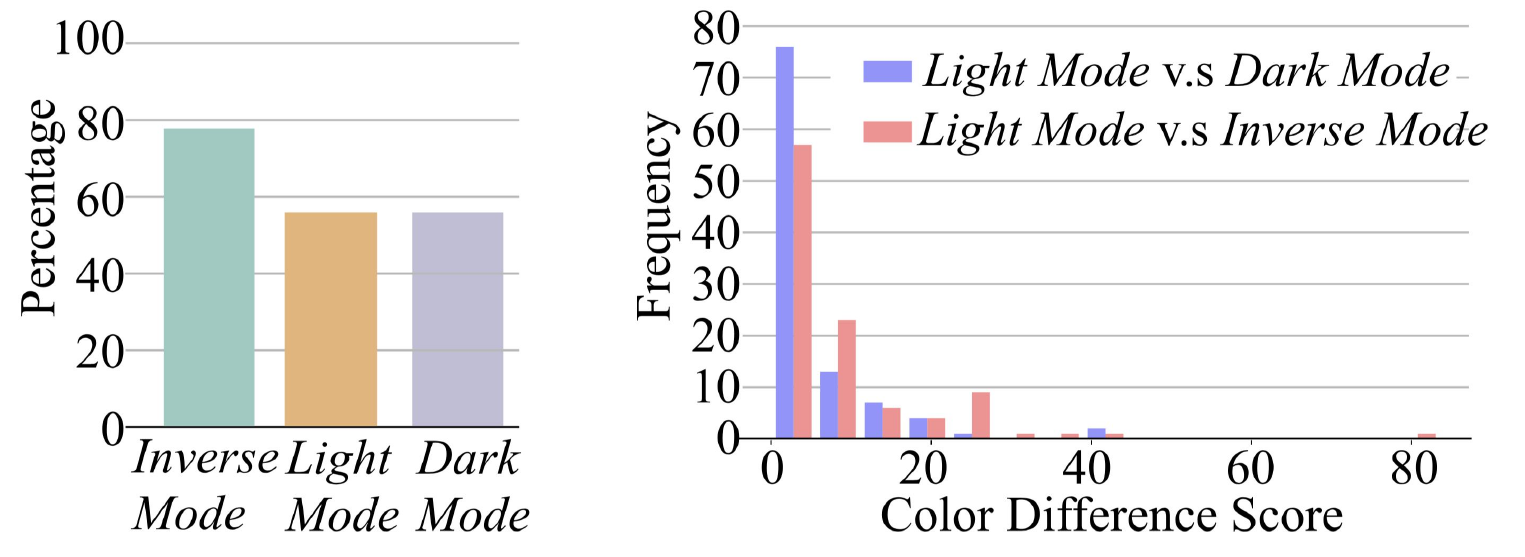}
    \caption{The evaluation results of visualization legibility and color difference. The percentage of charts meeting WCAG 2.1 contrast (left) and the histogram of color difference scores (right, binned by 5).}
    \label{fig:quantitative_evaluation_result}
\end{figure}

The evaluation revealed tradeoffs between legibility and color consistency (\cref{fig:quantitative_evaluation_result}). 
The inverse mode achieved the highest legibility, with 77.6\% of visualizations meeting WCAG standards compared to 67\% for both light and dark modes. 
However, the dark mode better preserved color semantics, showing more instances of low color difference scores (0-5 range) compared to the inverse mode.

The equal WCAG compliance rate between light and dark modes reveals a key limitation: our luminance contrast consistency approach maintains the original visualization's contrast ratios, meaning poor contrast in input visualizations leads to similar issues in dark mode. 
However, dark mode's superior color consistency demonstrates that our method effectively preserves semantic color relationships while adapting to dark backgrounds.

\subsection{User Study}

Following the system evalaution, we conducted a user study with 12 participants (4 females, 8 males, ages 24-31) to evaluate how users perceive and interpret dark mode visualizations generated by \toolName{}. 
All participants had normal or corrected-to-normal vision, no color blindness, and regularly used data visualization in their work. 
The study was approved by our university’s
Institutional Review Board.

\begin{figure}
    \centering
    \includegraphics[width=0.75\linewidth]{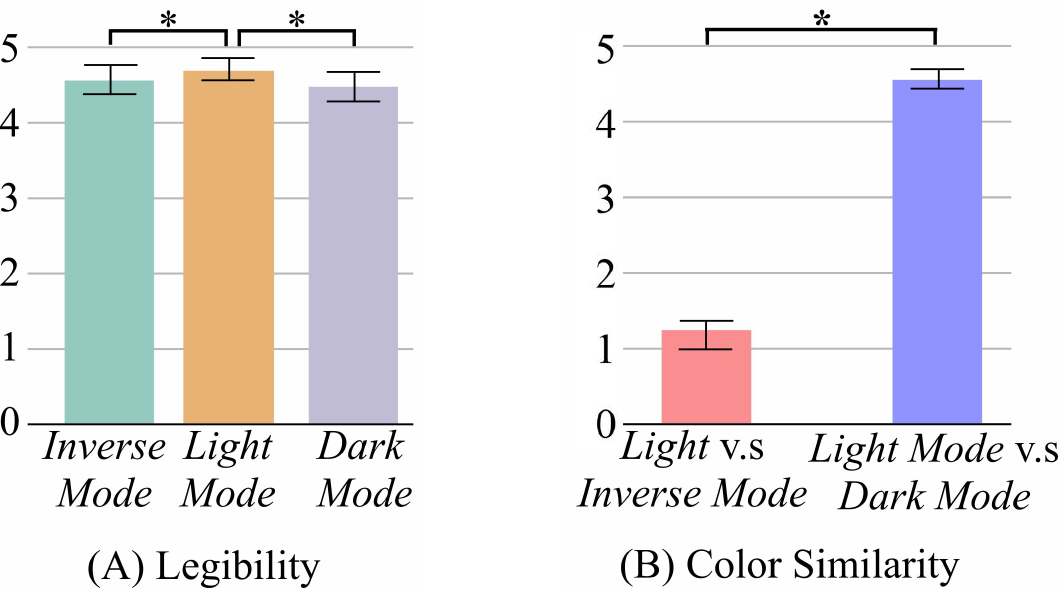}
    \caption{The results of legibility and color similarity in user study. The * indicates statistically significant differences between different conditions.}
    \label{fig:read_color_user_study}
\end{figure}

\subsubsection{Study Design}
We evaluated three visualization conditions: light mode, dark mode, and inverse mode. 
The study comprised three types of tasks. 
For legibility assessment, participants rated how clearly they could read single charts on a 7-point Likert scale (1=very difficult, 7=very easy). 
In the color similarity assessment, participants compared how similar dark mode and inverse mode visualizations were to their light mode versions (7-point scale, 1=very different, 7=very similar). 
For analytical tasks, participants performed identification and comparison tasks on multiple-view visualizations, following Munzner's task categorization~\cite{munzner2014visualization}. 
We measured accuracy, completion time, and visual fatigue (via a 10-question survey~\cite{Heuer1989RestPO}).

\begin{figure}
    \centering
    \includegraphics[width=1\linewidth]{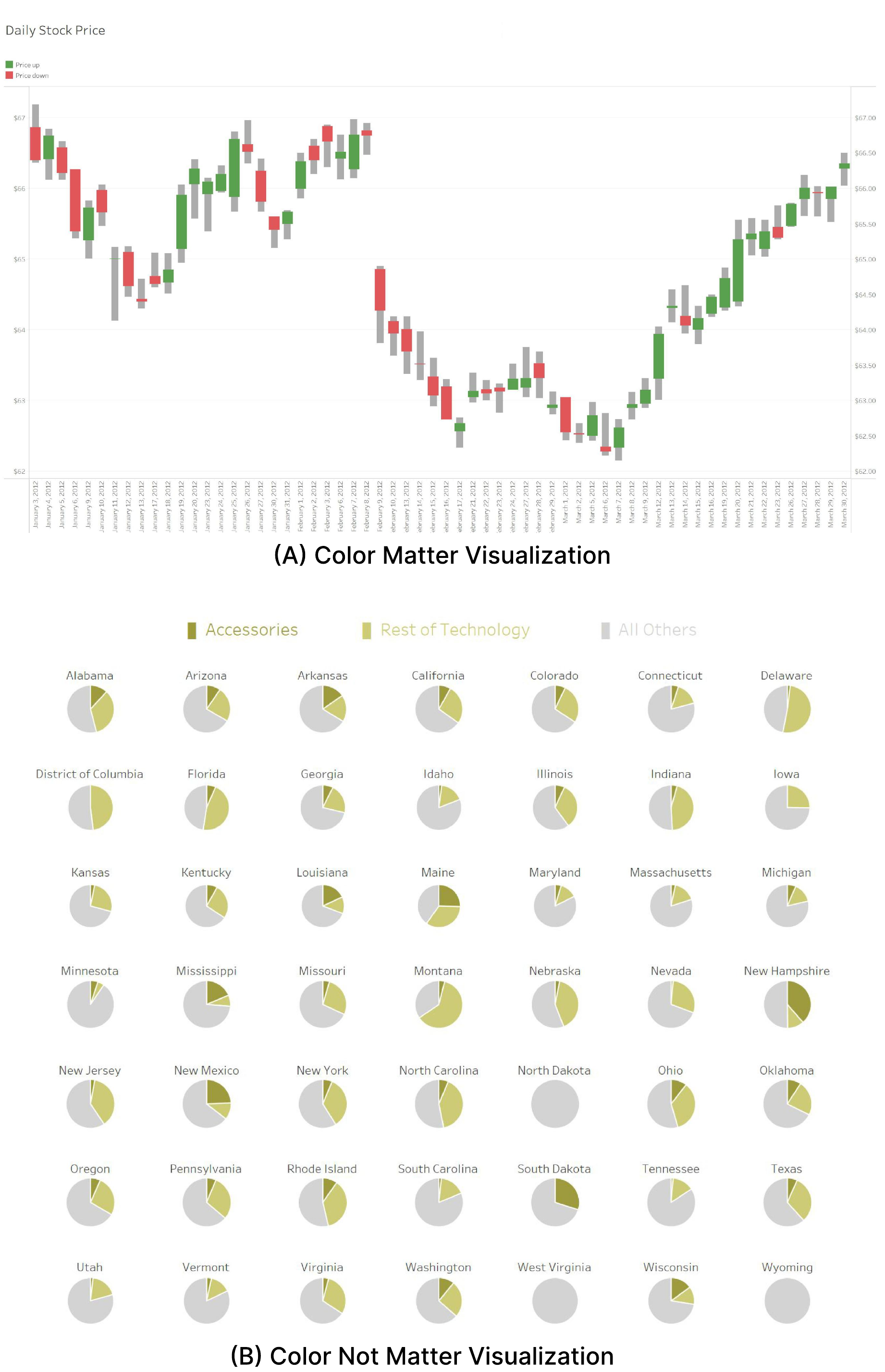}
    \caption{Examples of multiple-view visualization used in the user study. (A) A visualization displays a company's stock prices for one quarter. (B) A visualization showcases the contribution of three types of products to total sales in U.S. states.}
    \label{fig:stock_price}
\end{figure}

\begin{table}[t]
\small
\begin{tabular}{p{0.15\columnwidth}p{0.38\columnwidth}p{0.37\columnwidth}}
\toprule
\textbf{Vis.} & \textbf{Identification Task} & \textbf{Comparison Task} \\
\midrule
V1 & What date in the Coca-cola 2010 Q1 quarter did the stock price increase most? & 
What top 3 dates in the Coca-cola 2009 Q2 quarter did the stock price increase most? \\
\midrule
V2 & In which state did Rest of Technology account for 15\% of total sales? & 
Which three states have the highest portion of Rest of Technology make up for total sales? \\
\bottomrule
\end{tabular}
\caption{Sample questions for different tasks. V1 and V2 refer to the two visualizations shown in \cref{fig:stock_price}.}
\label{table:exp_sample_questions}
\end{table}

For the analytical tasks, participants performed value identification and comparison tasks on multiple-view visualizations. 
Each visualization was paired with two questions: one focused on identifying values and the other on comparing them. 
This framework was based on Tamara Munzner’s task categorization~\cite{munzner2014visualization}, which classifies analytical tasks into identification, comparison, and summary. 
The identification and comparison tasks (Table~\ref{table:exp_sample_questions}) were selected because they are fundamental analytic tasks and require users to engage intensively with the visualizations. 
Tasks were designed to evaluate user performance across the three visualization modes. 
Metrics included task accuracy, completion time, and a visual fatigue index derived from a 10-question survey~\cite{Heuer1989RestPO}, with higher scores indicating more fatigue.

\subsubsection{Materials}
We used two sets of visualizations. The first set included 16 single charts from the D3 Graph Gallery, covering a range of common visualization types (e.g., line charts, bar charts, pie charts) and complex types (e.g., choropleth maps, streamgraphs). The second set comprised 4 multiple-view visualizations from Tableau Public Gallery~\cite{tableauGallery}, divided into two categories (\cref{fig:stock_price}): visualizations where colors had conventional meanings (e.g., green for increase, red for decrease in stock price) and visualizations where colors had no inherent meaning.

\subsubsection{Procedure}
Participants completed tasks in a fixed task section sequence with randomized stimuli presentation. 
After completing a demographic questionnaire and tutorial, participants performed the legibility assessment, followed by the color similarity assessment, and then the analytical tasks with multiple-view visualizations. 
We used a balanced Latin square design to control for order effects, with 5-minute breaks between task sets. The participants received \$15 as compensation for their time.

\subsubsection{Results}

\begin{figure}
    \centering
    \includegraphics[width=1\linewidth]{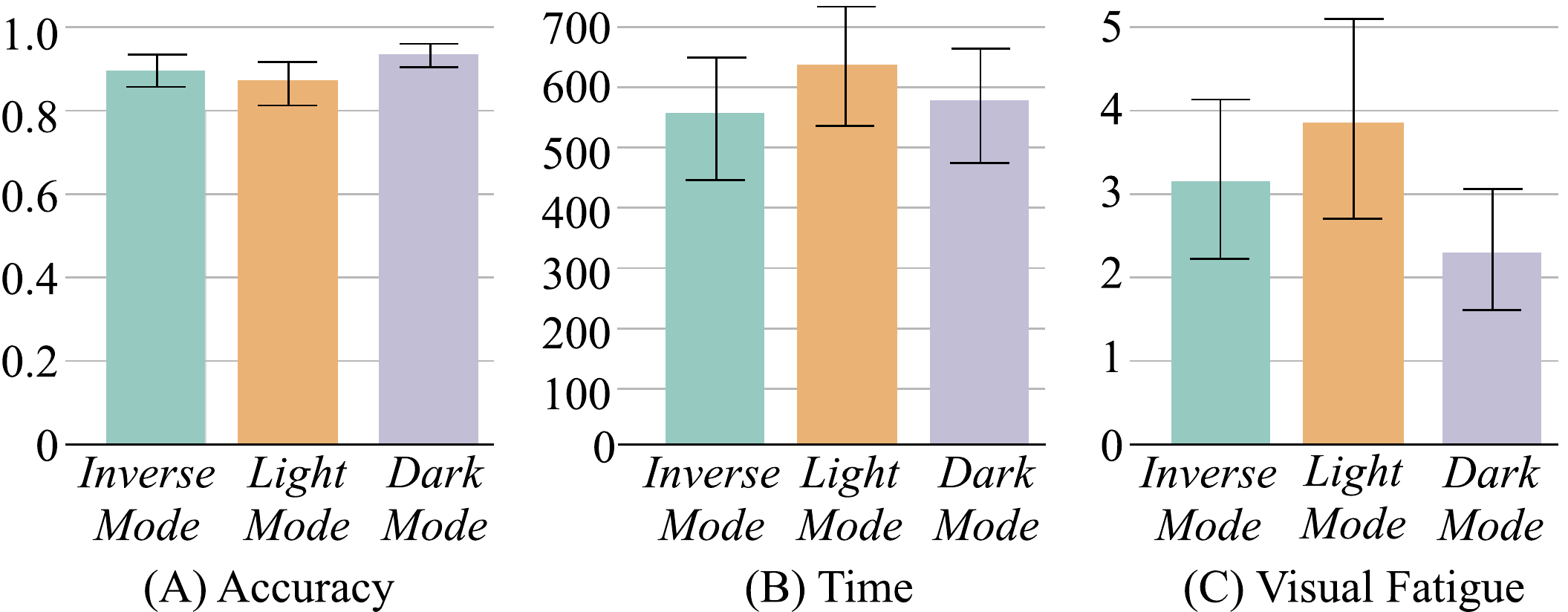}
    \caption{The results of accuracy, time, and visual fatigue Index in user study's analytical tasks. No statistically significant differences were observed across the three metrics.}    \label{fig:accuracy_time_fatigue_user_study}
\end{figure}

\begin{figure}
    \centering
    \includegraphics[width=1\linewidth]{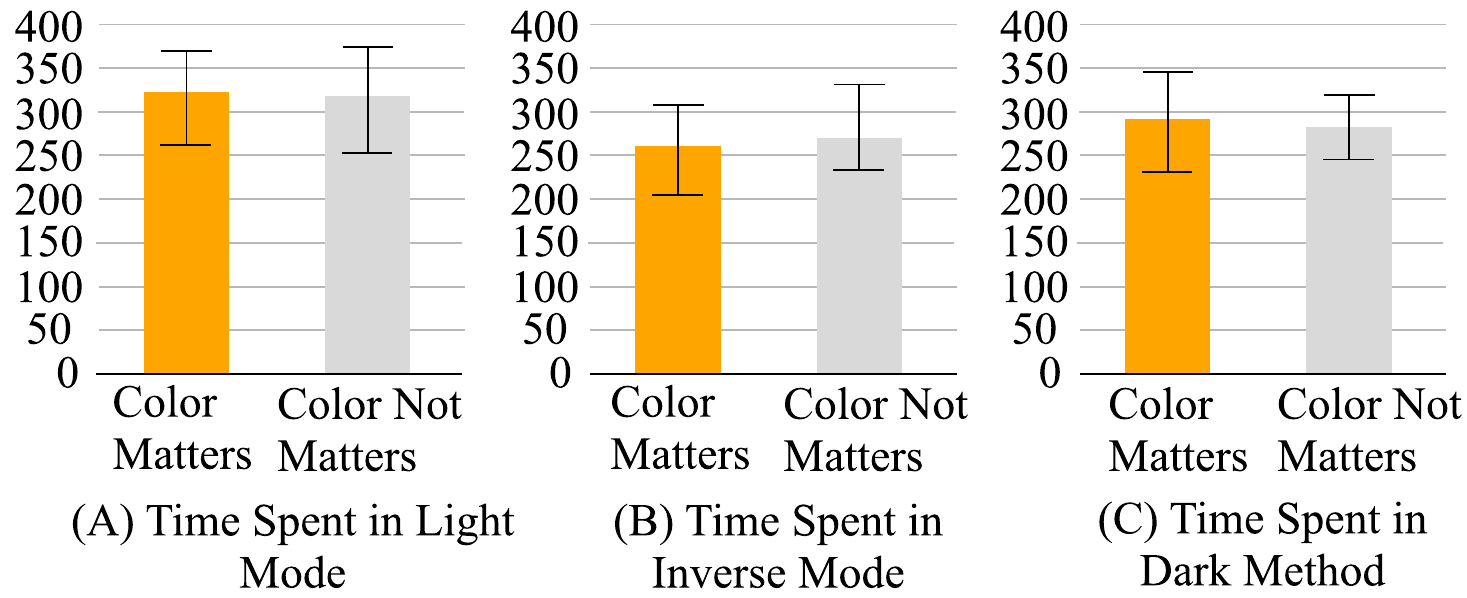}
    \caption{Time costs of tasks about \textit{Color Matters} v.s. \textit{Color Does Not Matter}. The figure compares the time (seconds) required to complete tasks across the three visualization conditions. No statistically significant differences were observed in any condition.}
\label{fig:time_color_matters_not_user_study}
\end{figure}

Depending on the distribution, we applied either a repeated measures t-test for normally distributed data or a repeated Wilcoxon test for non-normally distributed data. Our analysis revealed varying effects across three key dimensions: legibility, color similarity (\cref{fig:read_color_user_study}), and analytical performance (\cref{fig:accuracy_time_fatigue_user_study}). For legibility, while all three conditions achieved scores above the good legibility threshold of 4, light mode (M=4.651, $\sigma$=0.321) demonstrated significantly better legibility compared to both dark mode (M=4.344, $\sigma$=0.445, p=0.0156) and inverse mode (M=4.438, $\sigma$=0.438, p=0.0239), with no significant difference between dark mode and inverse mode. Regarding color similarity, participants perceived dark mode colors as significantly more similar to the original light mode (M=4.396, $\sigma$=0.382) compared to inverse mode colors (M=1.392, $\sigma$=0.489, p=0.000488). This suggests that our transformation method better preserves the original color relationships than simple color inversion. 

For analytical task performance, we found comparable effectiveness across all three modes (\cref{fig:accuracy_time_fatigue_user_study}). There were no significant differences in accuracy (dark mode: M=0.906, $\sigma$=0.0902; inverse: M=0.885, $\sigma$=0.108; light: M=0.865, $\sigma$=0.157), completion time (dark: M=560.18s, $\sigma$=134.99; light: M=638.70s, $\sigma$=161.44; inverse: M=539.20s, $\sigma$=154.81), or visual fatigue (dark: M=2.267, $\sigma$=1.289; inverse: M=3.116, $\sigma$=1.565; light: M=3.833, $\sigma$=2.332). Importantly, performance remained consistent regardless of whether colors carried semantic meaning (\cref{fig:time_color_matters_not_user_study}), indicating that our method maintains visualization effectiveness even when color interpretations are critical to the analysis task.

\section{Discussion}

\textbf{Strengths and Limitations.} 
The evaluation of \toolName{} demonstrates its effectiveness in transforming light mode visualizations while maintaining color semantics and legibility. However, several limitations emerged: 
\toolName{} treats all visualization elements equally, not distinguishing between text, axes, and graphical components that require different contrast levels according to WCAG 2.1 guidelines~\cite{w3ContentAccessibility}. 
This limitation, identified in expert feedback, often led to poor contrast and legibility in text and axes.
In future, a post-processing step to detect and invert text and axis colors could mitigate this issue.
Some color choices, particularly greens and yellows, were considered visually unappealing due to prioritizing legibility over aesthetics. 
Additionally, \toolName{} assumes similar color interactions in both modes and depends on the input visualization's contrast quality. 
The tool's relatively long execution time, while not critical for offline processing, could limit real-time applications.

\textbf{Applications and Future Directions.}
\toolName{} serves primarily as a design tool rather than an end-user solution. 
It is particularly valuable for mobile visualizations and critical data displays in varying lighting conditions, such as health monitoring or financial analysis dashboards. 
While currently limited to static images, future work could extend to SVG and interactive visualizations, enabling more precise element control through XML tags and dynamic color adaptation. 
Supporting color-blind users through specialized palette transformations would also enhance accessibility.
 
\textbf{Implementation Guidelines.} 
Two key parameters affect tool's performance. 
The cluster count ($k$) in color extraction significantly impacts transformation quality, \eg multi-view visualizations benefit from more clusters for better color palette extraction, while single view charts require fewer clusters for effective transformation. 
The weights of the three loss functions also require careful consideration. 
The Luminance Contrast Consistency weight ($\omega_0$ in Equation~\eqref{eq:object_function}) should be set at 1.0 to ensure readability matches the original visualization. 
The Color Consistency weights ($\omega_1$ in Equation~\eqref{eq:object_function}) should vary by visualization type, with continuous colormaps benefiting from lower weights (0.5) to preserve smooth transitions, while categorical data visualizations need higher weights (1.0-1.5) to maintain semantic meaning. 
The Adjacent Color Difference Consistency weight ($\omega_2$ in Equation~\eqref{eq:object_function}) should be at 1.0 to preserve relative differences between colors and maintain visual harmony. 
These parameters can be adjusted based on specific visualization needs and contexts. 
While our evaluation used controlled conditions with a black background, \toolName{}'s approach is device-independent and adaptable to various dark mode backgrounds.

\section{Conclusion and Future Work}

We present \toolName{}, an algorithm that converts light mode visualizations to dark mode while ensuring legibility, semantic relationships, and visual harmony. 
Utilizing dark mode UI design principles, our method employs a simulated annealing-based optimization focusing on luminance contrast consistency, color semantic consistency, and adjacent color difference consistency. 
Through case studies, expert interviews, and user studies, we show that \toolName{} effectively automates dark mode transformation, offering designers optimized starting points for refinement. 

In future work, we plan to broaden the usability of \toolName{} by extending support beyond static images to include scalable vector graphics (SVG), enabling finer control over individual elements. 
We also plan to adapt the method for mobile app interfaces, where dynamic context (e.g., changing light conditions) requires adaptive and lightweight transformations. 
Together, these directions will enhance the practicality of \toolName{} for real-world applications and improve accessibility across devices and platforms.
Further, with the increasing usage of Large Language Models (LLMs) for visualization tasks~\cite{fan2025well,wang2023llm4vis}, it will also be interesting to explore how LLMs can benefit automated color palette adaptation for dark mode visualizations.

\begin{acks}
This project is supported by the Ministry of Education,
Singapore, under its Academic Research Fund Tier 2 (Proposal ID:
T2EP20222-0049).
\end{acks}
\bibliographystyle{ACM-Reference-Format}
\bibliography{sample-base}

% \appendix









\end{document}